\documentclass[preprint,5p,times,twocolumn]{elsarticle}
\usepackage[utf8]{inputenc}
\usepackage[libertine]{newtxmath}

\usepackage{graphicx}                            
\usepackage{latexsym}                            
\usepackage{amsfonts}                            
\usepackage{amssymb}                             
\usepackage{amsmath}                             
\usepackage[mathscr]{eucal}                      
\usepackage{dcolumn}                             
\usepackage{theorem}                             
\usepackage{footnote}
\usepackage{enumitem}
\usepackage{bm}
\usepackage{color}
\usepackage{natbib}
\usepackage[english]{babel}
\usepackage[colorlinks,linkcolor=blue,anchorcolor=blue,citecolor=blue,urlcolor=blue,breaklinks=true]{hyperref}
\usepackage{physics}

\newcolumntype{C}[1]{>{\centering\arraybackslash}p{#1}}

\begin{document}

\title{Electromagnetic form factors for nucleons in short-range correlations}
\author{Wanli~Xing${}^{a}$,
Xuan-Gong~Wang${}^{a}$,
Anthony~W.~Thomas${}^{a}$\\[5mm]
\itshape{$^a$ CSSM, Department of Physics, University of Adelaide, Adelaide SA 5005, Australia}
}
\begin{abstract}
{Recent experimental studies have led to the suggestion that short-range correlations may be a major contributor to the nuclear EMC effect. This hypothesis requires that the structure function for nucleons involved in short-range correlations should be heavily suppressed compared to that of a free nucleon. Based on calculations performed within an AdS/QCD motivated, light-front quark-diquark model, we find that this large suppression of the nucleon structure function leads to a strong suppression of the nucleon elastic form factors.}
\end{abstract}
\maketitle

\section{Introduction}

 In discussing the nuclear EMC effect, we refer to the observation that the deep-inelastic structure functions differ significantly in the valence region for nuclei of different sizes~\cite{EuropeanMuon:1983wih, BCDMS:1987ovm, EuropeanMuon:1988lbf, Gomez:1993ri}. Following the original experiment by the European Muon Collaboration in 1983~\cite{EuropeanMuon:1983wih}, the effect has been thoroughly verified by numerous other experiments and it is widely accepted as evidence that the parton distributions are modified by the nuclear environment (see, for example, Refs.~\cite{Geesaman:1995yd, Norton:2003cb, Malace:2014uea, Rith:2014tma} for reviews). 

On the theoretical side, however, the mechanism underlying the EMC effect remains controversial, despite four decades of extensive research. 
Amongst the many proposals (for early examples see 
Refs.~\cite{Ericson:1983um,LlewellynSmith:1983vzz,Close:1983tn,Close:1984zm,Dunne:1985cn}) that aim to explain the effect, two schools of thought seem most promising. One is based on the modification of the bound nucleon structure resulting from the effects of the strong mean scalar and vector fields inside nuclei, a straightforward mechanism that has been shown to accurately describe the observed EMC 
effect~\cite{Thomas:1989vt,Saito:1992rm,Smith:2002ci,Mineo:2003vc,Cloet:2006bq}. The other explanation, which is the focus of this paper, suggests short-range correlations (SRCs) as the cause of the EMC 
effect~\cite{Frankfurt:1988nt,Weinstein:2010rt}. \\

Short range correlations describe the experimentally verified phenomenon that in a nucleus, 
nucleons sometimes form temporary pairs that scatter into a state of high relative momentum, naturally associated with a short distance scale~\cite{Hen:2016kwk,CLAS:2018yvt}. If the structure of a nucleon involved in an SRC were modified, the fact that leptons can at times scatter from such a nucleon would lead to a modification of the DIS cross section. Empirically, the probability of a nucleon being found in an SRC pair is strongly correlated with the magnitude of the EMC effect~\cite{CLAS:2019vsb}. This observation has led many to consider that SRC may even be the main driving-force behind the EMC effect. 

It has been realised that these two models give very different predictions regarding the polarised EMC effect. In particular, within the SRC-driven EMC model there should be essentially no polarised EMC effect~\cite{Thomas:2018kcx}. On the other hand, in the mean-field approach one expects to find significant polarised EMC effects for both quark~\cite{Cloet:2006bq,Cloet:2005rt} and gluon~\cite{Wang:2021elw} distributions. Hence, experiments that measure the polarised EMC effect would be extremely valuable in helping to resolve the debate. The proposed measurement of the polarised EMC effect in $^{7}$Li is expected to run in the next few years~\cite{jlab:2014}. The upcoming Electron-Ion Collider (EIC) will be a powerful tool to precisely probe gluonic aspects of the structure of nucleons and nuclei~\cite{Accardi:2012qut, AbdulKhalek:2021gbh}.

In the meantime, alternative means to gain some insight into the EMC effect are highly desired. 
Recently, it was shown that, in an $x$-rescaling model, the mass deficits of the SRC nucleons are insufficient to generate the observed nuclear EMC ratios~\cite{Wang:2022kwg}. Here we examine the effect on the electromagnetic form factors of  protons involved in SRC under the hypothesis that they drive the EMC effect.

In Sec.~\ref{sec:structure-function}, we briefly review the change in the nucleon structure functions proposed in the SRC model of the EMC effect. In Sec.~\ref{sec:form-factors} we first use quark-hadron duality to explore the qualitative expectation of the effect of the change in the structure functions of nucleons in SRC on the corresponding electromagnetic form factors. This is followed by a quantitative analysis within a di-quark model motivated by AdS/QCD. We present the numerical results in Sec.~\ref{sec:results}. Finally, in Sec.~\ref{sec:conclusion} we summarise our conclusions.

\section{Structure functions in SRC}
\label{sec:structure-function}
\begin{figure}[h!]
    \centering
    \includegraphics[width=0.49\textwidth]{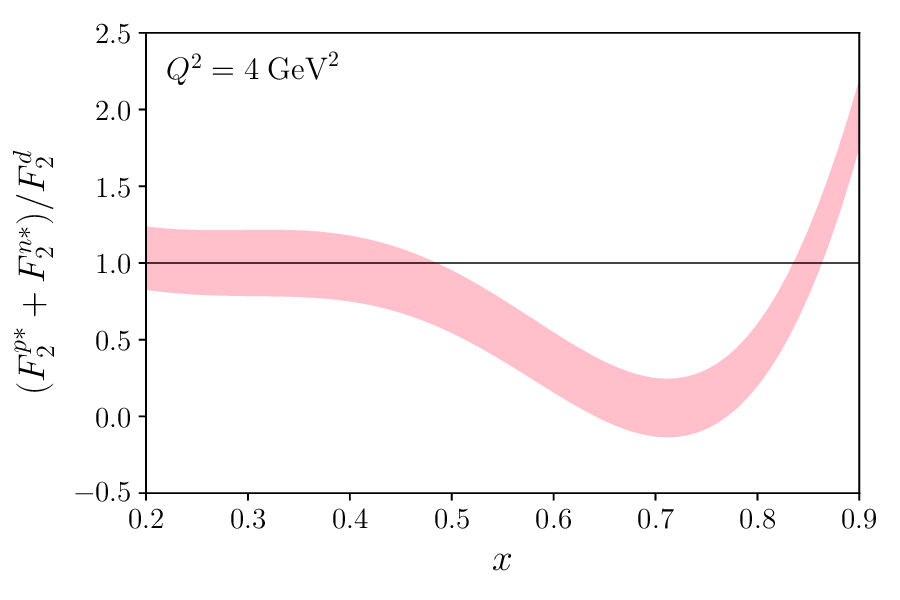}
    \caption{Ratio of the off-shell nucleon structure function in the SRC model to the deuteron structure function $(F_2^{p*}+F_2^{n*})/F_2^d$. This graph is similar to Fig.~2 of Ref.~\cite{Wang:2020uhj} but recalculated by taking $n^d_{\rm SRC}=0.04$. Note that for $x\lesssim0.7$, $F_2^d\approx F_2^p+F_2^d$, the free nucleon structure function. }
    \label{fig:SRCrange}
\end{figure}

The defining feature of the SRC approach is that, for a nucleus with mass number $A$, the modification to its nuclear structure function, $F_2^A$, is caused by proton-neutron ($np$) SRC pairs, of which there are $n_\text{SRC}^A$ in a nucleus. Nucleons not in an SRC pair are unmodified. That is~\cite{CLAS:2019vsb},
\begin{eqnarray}
    F_2^A & =& (Z-n_\text{SRC}^A)F_2^p+(N-n_\text{SRC}^A)F_2^n \nonumber \\ 
    &+&n_\text{SRC}^A (F_2^{p*}+F_2^{n*}). 
\end{eqnarray}
where $F_2^p, F_2^n$ are the unmodified free nucleon structure functions. $F_2^{p*}$ and $F_2^{n*}$ can then be interpreted as the structure functions of nucleons in an SRC pair. Of course, this approach is a little simplified in that the well known growth of the EMC ratio as $x \, \rightarrow \, 1$ caused by Fermi motion~\cite{Bickerstaff:1989ch} is neglected. This leads to the rise in the ratio at large $x$ shown in Fig.~\ref{fig:SRCrange}, which is not directly related to the change in the structure function caused by SRC.

The SRC model of the EMC effect for the deuteron was critically examined in Ref.~\cite{Wang:2020uhj}. An unusual feature of the SRC model, which was pointed out there, is that the ratio of the nucleon structure function in the SRC model against that of deuteron,
\begin{align}
    \frac{F_2^{p*}+F_2^{n*}}{F_2^d} & \approx \frac{F_2^{p*}+F_2^{n*}}{F_2^p+F_2^n},
\end{align}
has a very sharp drop-off, as indicated in Fig.~\ref{fig:SRCrange}. The approximation $F_2^d \approx F_2^p+F_2^n$ is justified because we will be interested in the region $x\lesssim0.7$, where the ratio $F_2^d/(F_2^p+F_2^n)$ is typically within a few percent of unity~\cite{Wang:2020uhj,Cocuzza:2021rfn}. So Fig.~\ref{fig:SRCrange} demonstrates that the nucleon structure function in the SRC model is vastly suppressed compared to that of a free nucleon. 

The structure function ratio in Fig.~\ref{fig:SRCrange} depends on the value of $n^d_{\rm SRC}$. It may drop below 0 at some point if we take $n^d_{\rm SRC} = 0.03$~\cite{Wang:2020uhj}. In large $x$ region, the ratio increases because of Fermi motion.
The crucial feature is the sharp drop-off of the structure function in the valence region between $x\sim0.5$ to $x\sim0.7$. This makes the off-shell structure function predicted by the SRC model drastically different from that in mean-field models. While unusual, this may not be problematic by itself since the individual nucleon structure function is not experimentally observable. However, one might well ask whether such a drastic change in the off-shell DIS structure function might have consequences in other physical observables, such as the form factors describing elastic scattering. A study in a two-component holographic model by Kim and Miller found relatively small effects~\cite{Kim:2022lng}. However, we find it interesting to explore the model dependence of that conclusion.

The main objective of this paper is thus to address the question of whether the suppression observed in the structure function shown in Fig.~\ref{fig:SRCrange} has an impact on the electromagnetic form factors for nucleons in SRC pairs.

\section{Form Factors in SRC}
\label{sec:form-factors}
In this section, we first provide a qualitative response to our central question based on quark-hadron duality~\cite{Melnitchouk:2005zr,DeWitt:2002er}. Then we perform a quantitative analysis using the AdS/QCD wave functions corresponding to the light front quark-diquark model introduced in Ref.~\cite{Maji:2016yqo}, which connects the structure functions and the form factors in a straightforward manner.

\subsection{Quark-hadron duality}
\label{sec:quark-hadron}
In a deep-inelastic scattering (DIS) experiment, let the hadron target have momentum $p$, and the incoming virtual photon have momentum $q$. The hadronic tensor is then
\begin{align}
    W^{\mu\nu} & = W_1(x,Q^2)\qty(-g^{\mu\nu} +\frac{q^\mu q^\nu}{q^2}) \nonumber \\
    & +\frac{W_2(x,Q^2)}{M^2}\qty(p^\mu-\frac{q\cdot p}{q^2}q^\mu)\qty(p^\nu-\frac{q\cdot p}{q^2}q^\nu).
\end{align}
In the Bjorken limit, the cross section may be expressed in terms of the DIS structure functions, for example,  $W_2(x,Q^2)\xrightarrow[]{\text{Bj lim}}W_2^\text{DIS}(x)$. We expect that suppression of the DIS structure function $W_2^\text{DIS}(x)$ might lead to suppressed elastic scattering form factors $F_{1,2}(Q^2)$. A quick, albeit qualitative, way to see this is based upon the concept of quark-hadron duality~\cite{Melnitchouk:2005zr,DeWitt:2002er}. This begins with the observation that for a structure function, say $\nu W_2(x,Q^2)$, where $\nu\equiv q^0$ in the laboratory frame, its integral
\begin{align}
    \int_{x_1}^{x_2}dx\,\nu W_2(x,Q^2) \, ,
\end{align}
is roughly constant for different values of $Q^2$. 
This is known as quark-hadron duality, and leads to the result that
\begin{align}
     \int_{1-\epsilon}^1 dx\, \nu W_2^\text{DIS}(x)\approx \int_{1-\epsilon}^{1} dx \,  \nu W_2(x,Q^2) \, ,
\end{align}
where the right-hand side is proportional to the elastic form factors. Indeed, for elastic scattering~\cite{Thomas:2001kw} 
\begin{equation}
W_2(\nu,Q^2) = \frac{G_E^2(Q^2) + \frac{Q^2}{4 M^2} G_M^2(Q^2)}{1+ \frac{Q^2}{4 M^2}} \delta(\nu - \frac{Q^2}{2M}) \, ,
\end{equation}
so that, at finite $Q^2$, $\nu W_2(x,Q^2)$ contains a spike in the region $x \, \rightarrow \, 1$  representing elastic scattering. A suppression in the DIS structure function $W_2^\text{DIS}(x)$ necessarily leads to a suppression in the elastic form factors, at lower $Q^2$ notably the electric form factor.

\subsection{Light front quark-diquark model with AdS/QCD wave functions}
\label{sec:quark-diquark}

Here we summarise the formalism introduced in Ref.~\cite{Maji:2016yqo}, which is used to investigate quantitatively the relationship between DIS structure functions and elastic form factors. 

In the quark model, the proton state is represented 
as $p = \ket{uud}$ and the neutron as $n = \ket{udd}$. In the quark-diquark model, two of the three valence quarks in a proton combine to form a diquark state, which can be a scalar with third component of isospin 0, or a vector with third component of isospin 0 or 1. The three possibilities give rise to states $\ket{u(ud)^0}\equiv \ket{uS^0}$, $\ket{u(ud)^0} \equiv \ket{uA^0}$ and $\ket{d(uu)^1}\equiv \ket{dA^1}$ respectively, where the superscripts refer to the third component of isospin. The whole the proton state is thus
\begin{align}
    \ket{P,\pm} = C_S\ket{uS^0}^\pm+C_V\ket{uA^0}^\pm + C_{VV}\ket{dA^1}^\pm \, ,
\end{align}
where the superscript $\pm$ refers to spin. The constants $C_S,C_V,C_{VV}$ cannot be calculated from the model and must be fitted to experimental data.  
We take their values from Ref.~\cite{Maji:2016yqo}, $C_S^2=1.3872, C_V^2=0.6128$ and $C_{VV}^2=1$. Note that we obtain the neutron state by interchanging $u \, \leftrightarrow \, d$ in the proton state. 

The Dirac and Pauli form factors can be decomposed into flavour form factors as $F_i^{p(n)} = e_u F_i^{u(d)} + e_d F_i^{d(u)}$ for $i=1,2$, and the flavour form factors themselves can be decomposed into form factors associated with the scalar and vector diquarks~\cite{Cates:2011pz}:
\begin{align}
    F_i^{(u)}(Q^2) &= C_S^2F_i^S(Q^2) + C_V^2F_i^{(V)}(Q^2), \\
    F_i^{(d)}(Q^2) &= C_{VV}^2F_i^{(VV)}(Q^2) \, . 
\end{align}
The form factors, $F_i$, assume very simple forms when calculated using light-front wavefunctions~\cite{Brodsky:1997de}. A defining feature of the light-front quark-diquark model used in 
Ref.~\cite{Maji:2016yqo} is its choice of a modified AdS/QCD wavefunction~\cite{Maji:2016yqo,Brodsky:2007hb,deTeramond:2011aml}. Here we simply present the final results:
\begin{align}
    F_1^{(S)}(Q^2)&=N_S^2R_1^{(u)}(Q^2), \\
    F_2^{(S)}(Q^2)&=N_S^2R_2^{(u)}(Q^2), \\
    F_1^{(V)}(Q^2)&=\qty(\frac{1}{3}N_0^{(u)2}+\frac{2}{3}N_1^{(u)2})R_1^{(u)}(Q^2), \\
    F_2^{(V)} (Q^2)& = -\frac{1}{3}N_0^{(u)2}R_2^{(u)}(Q^2), \\
    F_1^{(VV)}(Q^2)&=\qty(\frac{1}{3}N_0^{(d)2}+\frac{2}{3}N_1^{(d)2})R_1^{(d)}(Q^2), \\
    F_2^{(VV)} (Q^2)& = -\frac{1}{3}N_0^{(d)2}R_2^{(d)}(Q^2).
\end{align}
If we let $\nu=u,d$, then $R_1^{(\nu)}(Q^2)$ and $R_2^{(\nu)}(Q^2)$ are given by
\begin{align}
    R_1^{(\nu)}(Q^2) &= \int dx \, \bigg[T_1^{(\nu)}(x)\frac{(1-x)^2}{\delta^\nu} \nonumber \\ 
    & + T_2^{(\nu)}(x)\frac{(1-x)^4}{(\delta^\nu)^2}\frac{\kappa^2}{M^2\ln(1/x)} \nonumber \\
    & \times \bigg(1-\frac{\delta^\nu Q^2}{4\kappa^2}\ln(1/x)\bigg)\bigg]\exp[-\delta^\nu \frac{Q^2}{4\kappa^2}\ln(1/x)], \label{eqn:R1}\\
    R_2^{(\nu)}(Q^2) & = \int dx\ 2T_3^{(\nu)}(x)\frac{(1-x)^3}{\delta^\nu}\exp[-\delta^\nu\frac{Q^2}{4\kappa^2}\ln(1/x)] \, , 
    \label{eqn:R2}
\end{align}
where
\begin{align}
    T_1^{(\nu)}(x) & = x^{2a_1^\nu}(1-x)^{2b_1^\nu-1}, \label{eqn:T1}\\
    T_2^{(\nu)}(x) & = x^{2 a_2^\nu-2}(1-x)^{2b_2^\nu-1}, \label{eqn:T2}\\
    T_3^{(\nu)}(x) & = x^{a_1^\nu+a_2^\nu-1}(1-x)^{b_1^\nu+b_2^\nu-1} \, . 
    \label{eqn:T3}
\end{align}
Here $\kappa=0.4$\ GeV~\cite{Maji:2016yqo,Chakrabarti:2013dda} and the nucleon mass is set to be $M=0.94$ GeV. The remaining variables, $a_{1,2}^\nu, b_{1,2}^\nu$, and $\delta^\nu$ are the 10 free fitting parameters of the model.  

The normalisation conditions are determined by the values of $F_i$ at $Q^2=0$. For example, for the proton, $F_1^{(u)}(Q^2=0) = n_u = 2, F_1^{(d)}(Q^2=0)=n_d=1, F_2^{(u)}(Q^2=0)=\kappa_u=1.673, F_2^{(d)}(Q^2=0)=\kappa_d=-2.033$. This leads to normalisation $F_1^{(S)}(0) = 1, F_1^{(V)}(0)=1, F_1^{(VV)}(0)=1$. These conditions allow one to fix the normalisation factors~\cite{Maji:2016yqo}, $N_S=2.0191,N_0^{(u)}=3.2050, N_0^{(d)}=5.9423, N_1^{(u)}=0.9895,N_1^{(d)}=1.1616$. 

It is often more convenient to use the Sachs form factors of the nucleons ($i=p,n$), as they are directly related to the distribution of electric charge and magnetisation:
\begin{align}
    G_E^i(Q^2)&=F_1^i(Q^2)-\frac{Q^2}{4M^2}F_2^i(Q^2), \\
    G_M^i(Q^2)&=F_1^i(Q^2)+F_2^i(Q^2) \, . 
\end{align}

Tab.~\ref{table:freePar} gives the parameters that reproduce well the world data~\cite{Cates:2011pz, Diehl:2013xca} (the uncertainties in these parameters may be found in the original work).
\begin{table}
\renewcommand\arraystretch{1.3}
    \centering
    \begin{tabular}{C{1cm}|C{1.2cm} C{1.2cm} C{1.2cm} C{1.2cm} C{1.2cm}}\hline\hline
        $\nu$ & $a_1^\nu$ & $b_1^\nu$ & $a_2^\nu$ & $b_2^\nu$ & $\delta^\nu$ \\ \hline
        $u$ & 0.280 & 0.1716 & 0.84 & 0.2284 & 1.0 \\ 
        $d$ & 0.5850 & 0.7000 & 0.9434 & 0.64 & 1.0 \\ \hline
    \end{tabular}
    \caption{The free parameters at $\mu_0^2=0.098$ GeV$^2$.}
    \label{table:freePar}
\end{table}
%


The values of the parameters in Table~\ref{table:freePar} correspond to an unknown reference scale, $\mu_0$. 
To determine $\mu_0$, we must examine the unpolarised parton distribution functions (PDFs), determined by~\cite{Maji:2016yqo}
\begin{eqnarray}
\label{eq:f1}
f_1^{\nu} &=& N^{(\nu)} \bigg[\frac{1}{\delta^{\nu}} x^{2a_1^{\nu}} (1-x)^{2b_1^{\nu}+1} \nonumber \\
&& + x^{2a_2^{\nu}-2} (1-x)^{2b_2^{\nu}+3}
\frac{\kappa^2}{(\delta^{\nu})^2 M^2 \ln(1/x)}\bigg]\ ,
\end{eqnarray}
where $\nu=u, d$ and the overall constants are
\begin{eqnarray}
\label{eq:Nu-Nd}
N^{(u)} &=& C^2_S N^2_S + C^2_V \left( \frac{1}{3} N_0^{(u)2} + \frac{2}{3} N_1^{(u)2} \right)\, ,\nonumber\\
N^{(d)} &=& C^2_{VV} \left( \frac{1}{3} N_0^{(d)2} + \frac{2}{3} N_1^{(d)2} \right)\, .
\end{eqnarray}
We optimise the initial scale to be $\mu_0^2 = 0.098$ GeV$^2$ such that the parton distribution $x(f_1^u + f_1^d)$ best reproduces the experimental data~\cite{Maji:2016yqo,Broniowski:2007si} after next-to-leading order (NLO) evolution~\cite{Bertone:2013vaa} to $Q^2=4\ {\rm GeV}^2$. The structure function $F_2^p + F_2^n$ for free nucleons is shown by the solid curve in Fig.~\ref{fig:SRCSF}.

\section{Results}
\label{sec:results}
\begin{figure}[h!]
\includegraphics[width=0.49\textwidth]{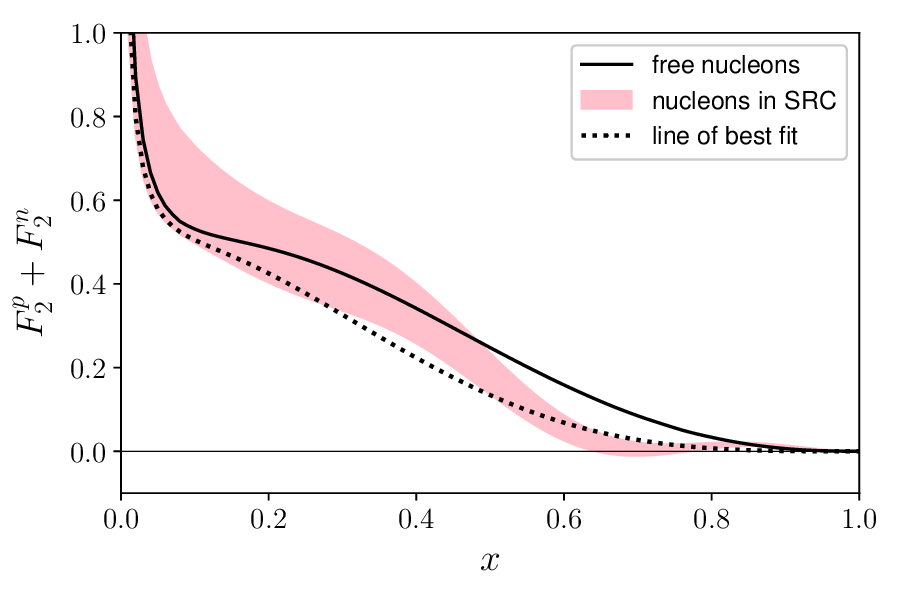}
\caption{The structure function $F_2^p+F_2^n$ at $Q^2=4$ GeV$^2$ for free nucleons (black solid line) and SRC nucleons (red band). The black dotted curve shows the fit obtained here after evolving the initial PDFs with the modified parameters given in Tab.~\ref{table:freePar2}.}
\label{fig:SRCSF}
\end{figure}

Recall from the discussion surrounding Fig.~\ref{fig:SRCrange} that the SRC-driven EMC model is characterised by a drastic suppression of the structure function in the region $0.5<x<0.7$, and our aim is to determine the effect of such a suppression on the elastic form factors. To this end, we multiply the solid line in Fig.~\ref{fig:SRCSF} by the ratio in Fig.~\ref{fig:SRCrange}, and the resulting structure function of nucleons in an SRC pair is shown by the red band in Fig.~\ref{fig:SRCSF}. We then modify the reference free parameters in Tab.~\ref{table:freePar} to generate a new distribution function which approximates the red band. Finally, we use these new parameters to calculate the form factors in the SRC model. 

 At the initial scale $\mu_0^2$, the parameters $\delta^{\nu}$ are always equal to 1 for both the $u$ and $d$ quarks~\cite{Maji:2016yqo}. Despite having 8 free parameters, achieving the desired sharp fall-off at mid-$x$ proves difficult. One challenge is that all free parameters enter as powers of $x$ or $(1-x)$ in Eqs.~\eqref{eqn:R1} to \eqref{eqn:T3}, and such functions do not increase or decrease rapidly in the mid-$x$ region when all free parameters are assumed to be between 0 and 1, as is the case in Ref.~\cite{Maji:2016yqo}. 

The dotted line in Fig.~\ref{fig:SRCSF} shows a typical result of the search for parameters that best reproduce the SRC structure function.
The modified parameters at the initial scale $\mu_0^2$ are given in Tab.~\ref{table:freePar2}.
\begin{table}[h!]
\renewcommand\arraystretch{1.3}
    \centering
    \begin{tabular}{C{1cm}|C{1.2cm} C{1.2cm} C{1.2cm} C{1.2cm} C{1.2cm}} \hline\hline
        $\nu$ & $a_1^\nu$ & $b_1^\nu$ & $a_2^\nu$ & $b_2^\nu$ & $\delta^\nu$ \\ \hline
        $u$ & 0.12 & 1.0 & 0.92 & 0.80 & 1.0 \\ 
        $d$ & 0.20 & 0.75 & 0.90 & 0.50 & 1.0 \\ \hline 
    \end{tabular}
    \caption{The parameters at $\mu_0^2=0.098$ GeV$^2$, which have been optimised to best fit the parton distributions of nucleons in SRC at $Q^2 = 4\ {\rm GeV}^2$.}
    \label{table:freePar2}
\end{table}

The light-front quark-diquark model in 
Ref.~\cite{Maji:2016yqo} is meant to incorporate the  dependence on the scale, $\mu$, made possible by making all of the free parameters, 
$\{a_i^\nu,b_i^\nu,\delta^\nu\}$, $\mu$-dependent. Analytic expressions at arbitrary $\mu$, chosen to reproduce the result of QCD evolution, 
$\{a_i^\nu,b_i^\mu,\delta^\nu\}(\mu)$, are given in Ref.~\cite{Maji:2016yqo}. However, since these analytic expressions for the scale dependence are obtained by fitting experimental data, one does not expect them to still be valid in the SRC case. Hence, instead of using the analytic expressions at $\mu$, we start with $\mu_0$ and numerically evolve our results to higher $\mu$. 

\begin{figure*}
  \begin{minipage}[b]{0.49\linewidth}
    \includegraphics[width=\linewidth]{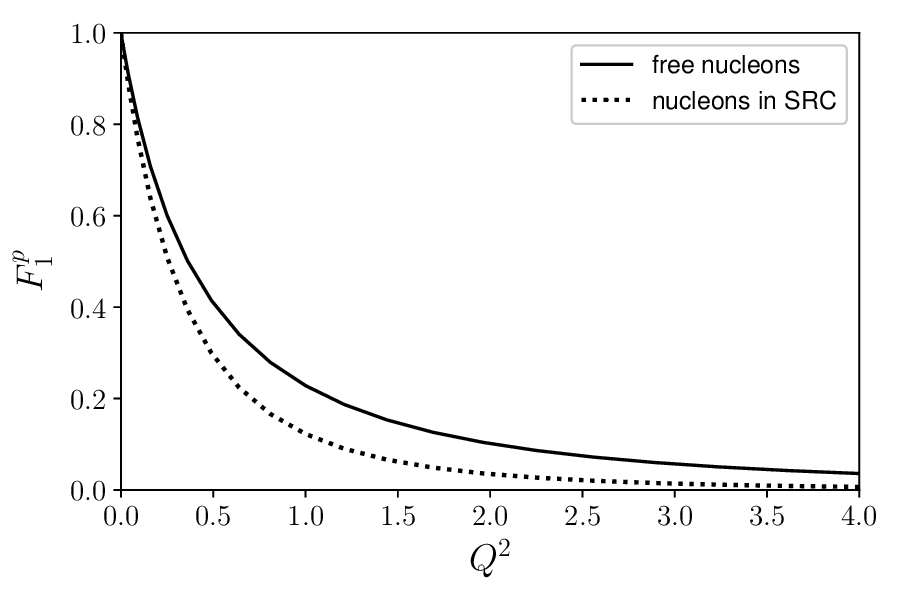} 
    \caption{Dirac form factor for the proton.} 
    \label{fig:F1p}
  \end{minipage} 
  \begin{minipage}[b]{0.49\linewidth}
    \includegraphics[width=\linewidth]{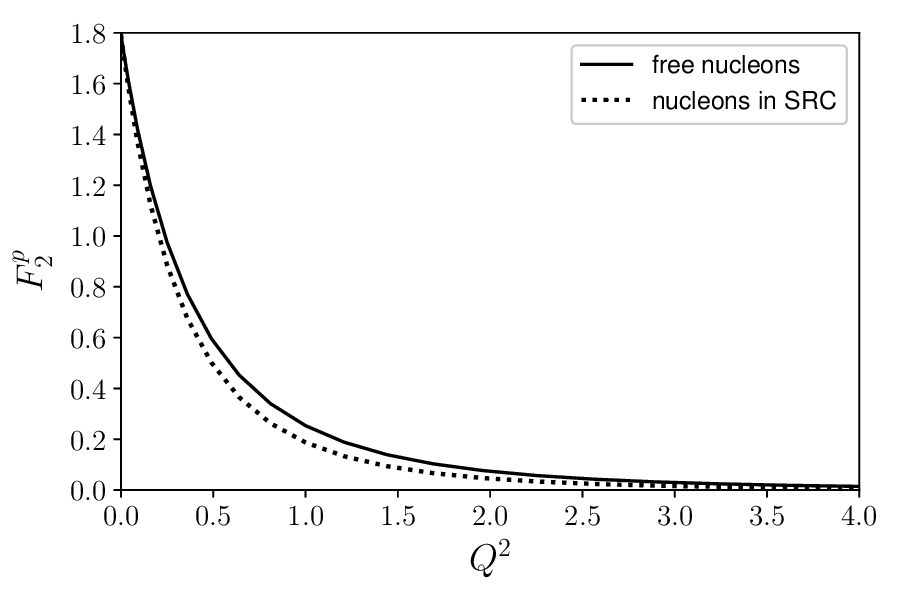}  
    \caption{Pauli form factor for the proton.}
    \label{fig:F2p}
  \end{minipage} 
  \begin{minipage}[b]{0.49\linewidth}
    \includegraphics[width=\linewidth]{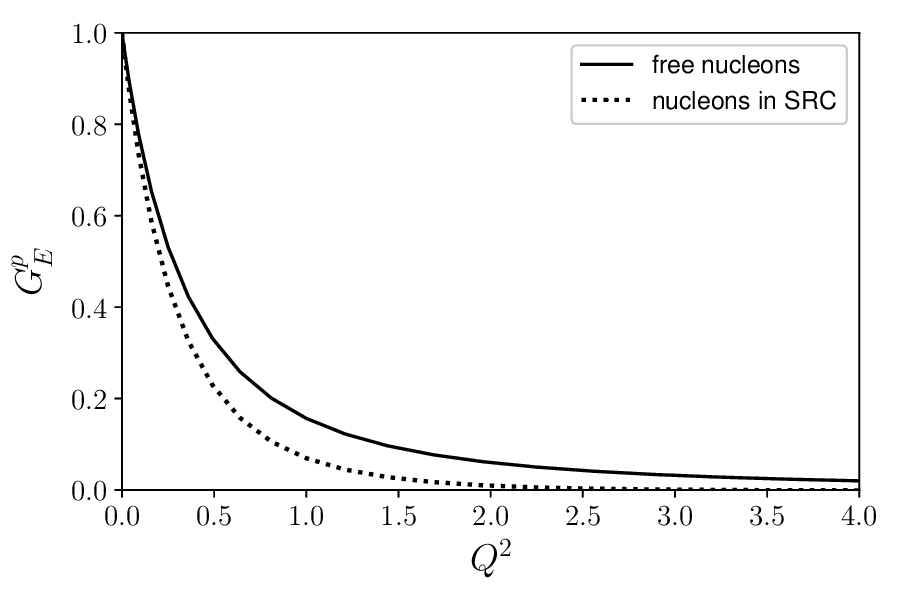} 
    \caption{Electric form factor for the proton.}
    \label{fig:GpE} 
  \end{minipage}
  \hfill
  \begin{minipage}[b]{0.49\linewidth}
    \includegraphics[width=\linewidth]{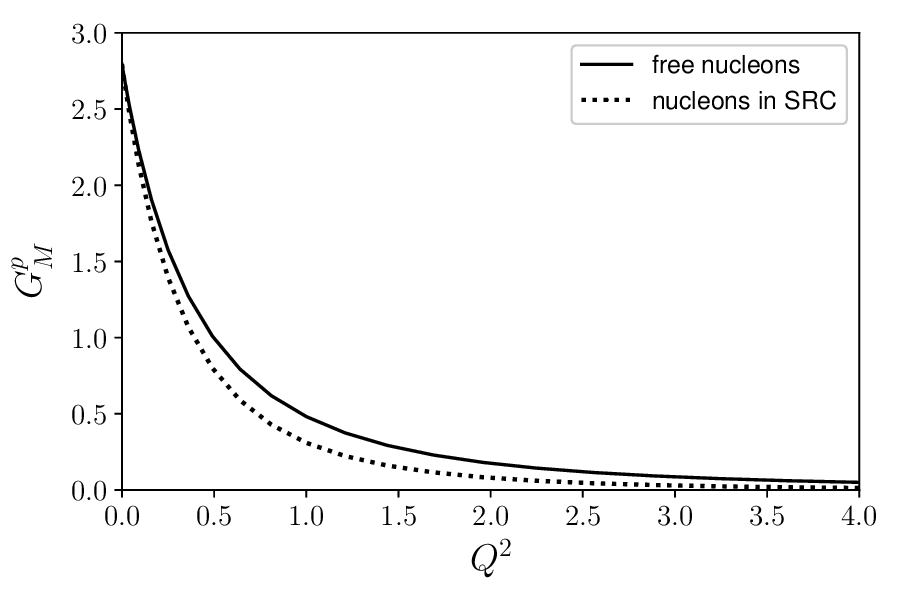} 
    \caption{Magnetic form factor for the proton.}
    \label{fig:GpM} 
  \end{minipage} 
\end{figure*}

In Figs.~\ref{fig:F1p} to \ref{fig:GpM}, we show the various form factors of free nucleons and nucleons in SRC, based on the parameters in Tab.~\ref{table:freePar} and Tab.~\ref{table:freePar2}, respectively.
We see that all of the proton elastic form factors in the SRC model are reduced away from $Q^2=0$. This suppression is most dramatic for the Dirac and the electric form factors, while the Pauli form factor and the magnetic form factor are less affected.

These changes in the elastic form factors would have a significant effect on the cross section for quasi-elastic scattering from a nucleus. In particular, in the case where a proton involved in an SRC is struck,  the longitudinal cross section would be reduced dramatically, for example by a factor of 4 at $Q^2 = 1$ GeV$^2$. However, the original investigation of nucleons in SRC, and especially the role of the tensor force, was carried out using quasi-elastic scattering. In this way the number of pairs in SRC was mapped out as a function of mass number~\cite{CLAS:2018yvt}. The large reduction in the quasi-elastic scattering cross section found here requires that the number of pairs in SRC deduced from the data would increase dramatically. This would lead to the conclusion that almost all the nucleons in a heavy nucleus must be involved in SRC, rather than the 20-30\% reported on the basis of the analysis using the free elastic form factors.

\section{Conclusion}
\label{sec:conclusion}
Within a model based upon AdS/QCD~\cite{Maji:2016yqo}, we have investigated the implications of the proposed suppression in the off-shell nucleon structure function for nucleons in SRC. We showed that this suppression in the deep inelastic structure functions leads to a sizeable suppression of the elastic form factors for nucleons in SRC. At least within this model, this suggests an inconsistency in the SRC model of the EMC effect. In particular, the suppression of the quasi-elastic scattering cross section for protons in SRC would imply that the number of neutron-proton pairs involved in SRC could have been underestimated by a large factor.

Our findings are supported both by qualitative arguments based on quark-hadron duality and a quantitative analysis using a light-front quark-diquark model. Given the potential importance of this result for the interpretation of the EMC effect, further investigation into the relationship between the off-shell nucleon structure function and the elastic form factors is urgently needed. 

\section*{Acknowledgement}
We are pleased to acknowledge helpful discussions with W. Melnitchouk.


\end{document}